\documentclass[aps,prl,reprint,groupedaddress,twocolumn,superscriptaddress]{revtex4-2}
\usepackage{graphicx}
\usepackage{hyperref}
\usepackage{subfigure}
\graphicspath{{figures/}}

\newcommand{\ket}[1]{\left| #1 \right\rangle}

\begin{document}
\title{
In-situ ac Stark shift Detection in Light Storage Spectroscopy
}
\author{D. Palani}
\address{Albert-Ludwigs-Universität Freiburg, Physikalisches Institut, 79104 Freiburg, Germany}
\author{D. Hoenig}
\address{Albert-Ludwigs-Universität Freiburg, Physikalisches Institut, 79104 Freiburg, Germany}
\author{L. Karpa}
\email[]{karpa@iqo.uni-hannover.de}
\address{Albert-Ludwigs-Universität Freiburg, Physikalisches Institut, 79104 Freiburg, Germany}
\address{Leibniz Universität Hannover, Institut für Quantenoptik, 30167 Hannover, Germany}
\date{\today}
\begin{abstract}
We report on a method for measuring ac Stark shifts observed in stored light experiments while simultaneously determining the energetic splitting between the electronic ground states involved in the two-photon transition. To this end we make use of the frequency matching effect in light storage spectroscopy. We find a linear dependence on the intensity of the control field applied during the retrieval phase of the experiment. At the same time, we observe that the light shift is insensitive to the intensity of the signal field which is in contrast to continuously operated schemes using electromagnetically induced transparency (EIT) or coherent population trapping (CPT). Our results may be of importance for future light storage-based precision measurements with EIT and CPT-type devices where, in contrast to schemes using continuous exposure to optical fields, the impact of intensity fluctuations from the signal field can be suppressed.
\end{abstract}
\maketitle
The coherent interaction of optical fields with multilevel systems gives rise to interesting phenomena such as Coherent Population Trapping (CPT) \cite{Alzetta1976,Arimondo1996} and EIT \cite{Harris1990,Fleischhauer2005}. These closely related effects arise from the emergence of non-interacting states comprised of coherent superpositions of electronic groundstates which in the ideal case have no contribution of the excited state. The former are responsible for the occurrence of narrow transmission peaks (EIT) \cite{Brandt1997} in otherwise opaque media and the associated drastically reduced group velocities known as slow light \cite{Hau1999,Kash1999,Budker1999}. These phenomena have fostered applications in magnetometry \cite{Scully1992,Katsoprinakis2006,Vanier2005} and 
atomic clocks \cite{Vanier2005} where the ac Stark shift induced by all optical fields on the energy levels of the interacting media is a major concern met with continuously developed elaborate measurement schemes to mitigate its impact \cite{Vanier2005,Ludlow2015a,ZanonWillette2018}.

Such experiments typically involve the coupling of two resonant optical fields, a strong control field, and a weaker signal field, to effective $ \Lambda $-type three-level atomic systems. Dynamic manipulation of the properties of the light fields allows for a reduction of the associated group velocity to zero and subsequent reversal of this process at a later time, a procedure referred to as light storage \cite{Fleischhauer2000,Liu2001,Phillips2001}.  It was shown that upon retrieval of a previously stored light pulse, the difference frequency between the signal and control fields matches the energy splitting of the ground states in the probed three-level system \cite{Karpa2009}. This effect known as frequency matching which can be understood within the framework of the polariton picture of EIT occurs regardless of the initial two-photon detuning and does not directly depend on the width of the transmission peaks making it potentially interesting as an alternative method for measuring magnetic fields.

Here, we demonstrate a method for measuring the energy splitting between two electronic ground states in a three-level system, including the energy shift stemming from the interaction with the optical fields. We achieve this by utilizing the effect of frequency matching in light storage spectroscopy. While measurement methods exploiting the narrow two-photon resonances arising from electromagnetically induced transparency (EIT) and coherent population trapping (CPT) are sensitive to intensity fluctuations of all involved optical fields, we show that the observed light shift in our method is determined by the properties of the control field only. That is, we demonstrate that the retrieved optical field does not cause any additional light shifts. This has implications for light storage-based sensing schemes and quantum memories since the evolution of the underlying spin coherence during the storage period is decoupled from any intensity fluctuations of the input pulses.\\

\begin{figure}[h!!!]
\centering
\includegraphics[width = 0.4 \textwidth]{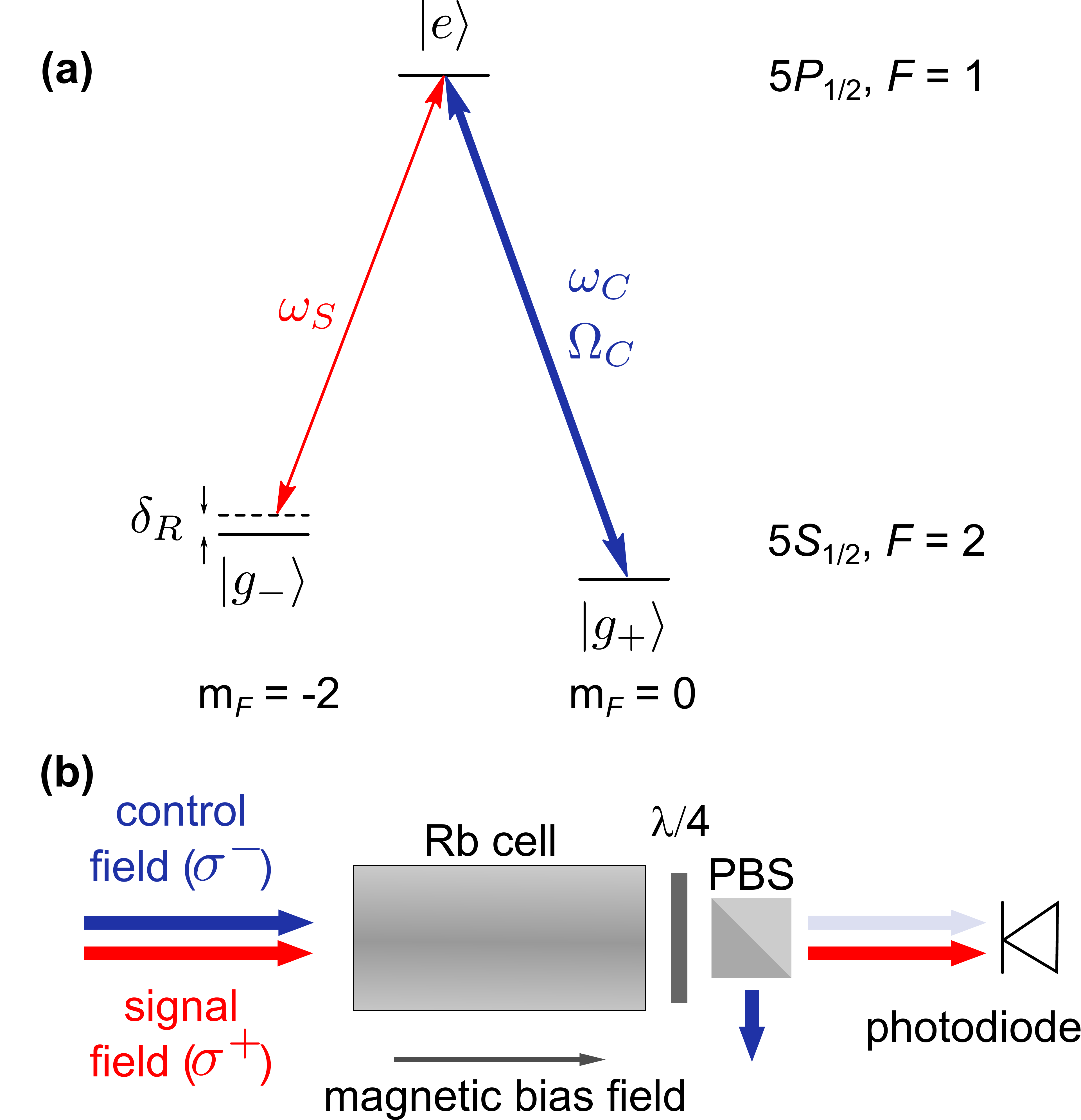}
\caption{
(a) Simplified level scheme of the $^{87}\text{Rb} $ D1 line. The electronic states comprising the idealized three-level $ \Lambda $ system are addressed by two optical fields, a circularly polarized $ (\sigma^-) $ control field with an optical frequency of $ \omega_C $ and a Rabi frequency $ \Omega_C $, and a $ \sigma^+ $-polarized signal field with a frequency of $ \omega_S $. The two-photon Raman detuning is denoted as $ \delta_R $. (b) Schematic of the experimental setup. The rubidium buffer gas cell is centered within a solenoid and enclosed in magnetic shielding (not shown). The polarizing beam splitter and a quarter-wave plate are denoted as PBS and $ \lambda/4 $, respectively.
}
\label{fig:setup}
\end{figure}
The setup used for our experiments is similar to a previously described apparatus \cite{Karpa2008,Karpa2009,Karpa2010,Djokic2015}.
We use a $ 50~\text{mm} $ long rubidium cell containing 20 torr neon acting as a buffer gas. The temperature of the buffer gas cell is actively stabilized to approximately $T \approx 90^\circ \text{C}$ with a relative uncertainty of $\Delta T = \pm 0.02~\text{K}$. The cell is positioned in the center of a solenoid (length: $L \approx 0.34 ~ \text{m} $; radius: $R \approx 0.05 ~ \text{m}$) providing an approximately homogeneous magnetic bias field $ \vec{B}_0 $ aligned with the propagation direction of the optical fields to control the splitting of the ground-state Zeeman sublevels denoted as $\ket{g_-}$ and $ \ket{g_+}$ in Fig. \ref{fig:setup}(a). The solenoid is enclosed in three layers of $ \mu $-metal shielding in order to isolate the experiment from ambient and stray magnetic fields. An external cavity diode laser serves as a source for both the control and signal field. Its optical frequency is stabilized to the $F = 2 \rightarrow F'= 1 $ transition of the rubidium D1 line near 795 nm by means of Doppler-free spectroscopy. A polarizing beam splitter (PBS) divides the laser beam into two fields, the control and signal field, which then pass independent acousto-optic modulators (AOM) to allow for precise control over their respective intensity and frequency. Subsequently, they are spatially overlapped at a second PBS and sent through a polarization maintaining single-mode optical fiber, ensuring spatial mode matching of the two fields. After exiting the fiber both fields are collimated to a beam diameter of $ 0.9~ \text{mm} $, pass through a $\lambda /4 $-plate and enter the cell with opposite circular polarisations.
After traversing the buffer gas cell, the light fields' polarizations are again converted to linear by using another $\lambda /4 $-plate. This allows us to remove the control field on a PBS and to detect the signal field intensity on a photodiode. Adjustment of this $\lambda /4 $-plate in combination with the PBS allows for controlling the mixture of control and signal light reaching the photodetector. %
\begin{figure}[h!!!]
\centering
\includegraphics[width = 0.5 \textwidth]{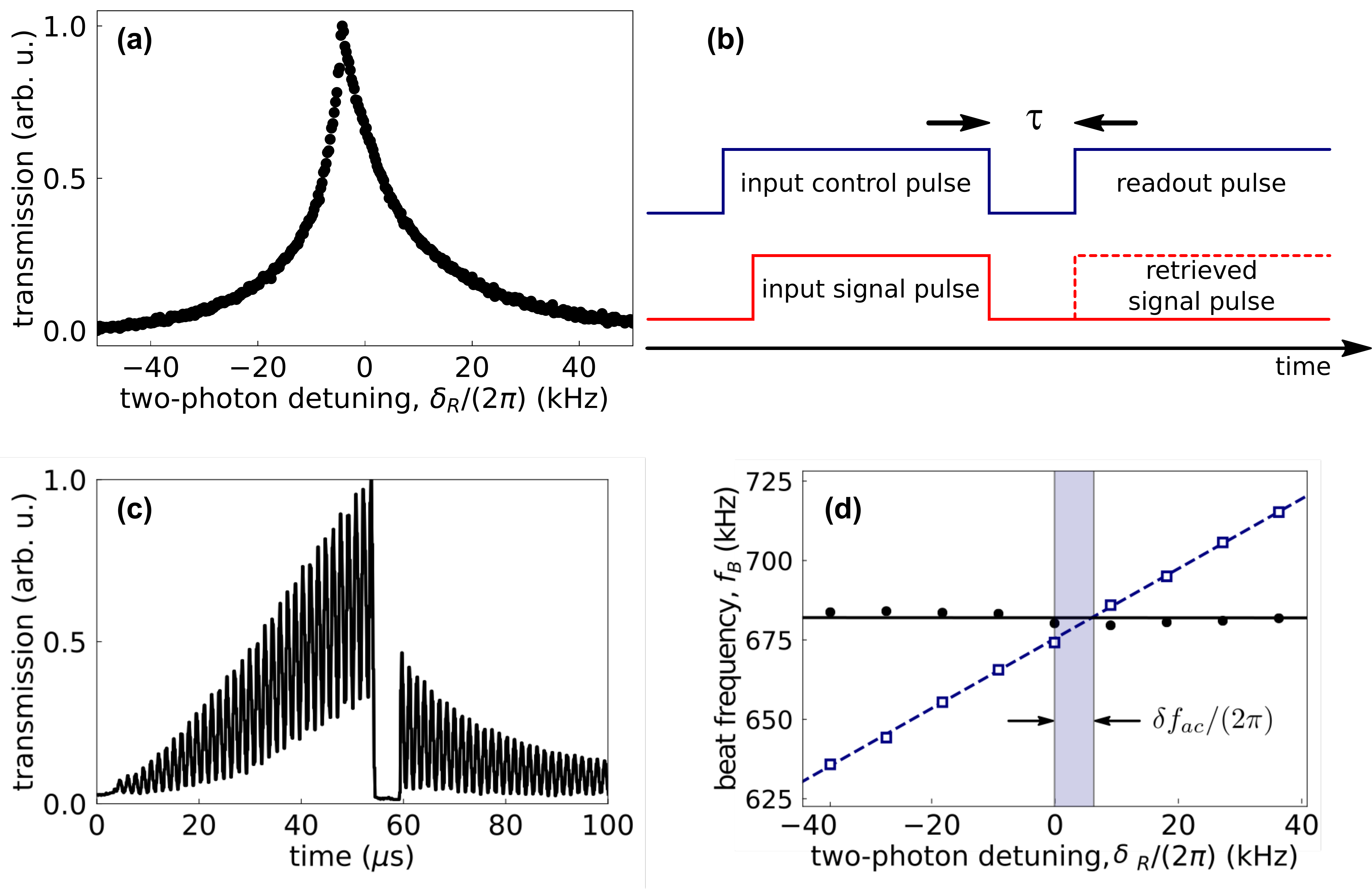}
\caption{
(a) Experimental transmission spectrum of a dark resonance with a width of approximately $ 20~\text{kHz} $ (full width at half maximum). The transmission of the signal field as a function of $ \delta_R $ at a fixed magnetic bias field $ B_0 $ shows a resonance at $ \delta_R = 0 $. 
(b) Pulse sequence for light storage experiments. We start the sequence by illuminating the rubidium atoms with a control field of a chosen intensity $ I_C $ to prepare the ensemble in the state $\ket{g_-}$. The signal field with a duration of $\sim 50~\mu$s and the control field are switched off simultaneously. After a storage period $ \tau \approx 5 ~ \mu$s, the control field is turned on again, triggering the retrieval of the signal field. (c) Light storage experiment recorded with a photodiode (average over 10 consecutive realizations). The rapid oscillations arise from beating of the input (for $ t < 55~\mu \text{s} $) and the retrieved signal pulses ($ t > 60~\mu \text{s} $) with the control field. In both cases, the corresponding beating frequency $ f_B $ is extracted by fitting a sinusoidally modulated function to the data. 
(d) Typical result of a light storage spectroscopy experiment. At fixed $ B_0 $, $ \delta_R $ is varied within the EIT window. The extracted values of $ f_B $ for the input signal and control fields (blue hollow squares) as well as the retrieved signal and readout control fields (black circles) are shown with linear fits to the data (blue dashed line for input pulses; black solid line for readout/retrieved pulses). The beat frequency of the pulses after retrieval shows the characteristic frequency matching, i.e. $ f_B $ is constant within experimental uncertainties. The beat notes of the input and the readout fields intersect at $ \delta_R = \delta f_{ac} $ due to a differential ac Stark shift between the states  $ \ket{F=2, ~ m_F = -2} $ and $ \ket{F=2, ~ m_F = 0} $ ($\ket{g_-}$ and $\ket{g_+}$ in Fig. \ref{fig:setup}(a)) induced by the control field.
}
\label{fig:schematic}
\end{figure}
\begin{figure}[h!!!]
\centering
\includegraphics[width = 0.45 \textwidth]{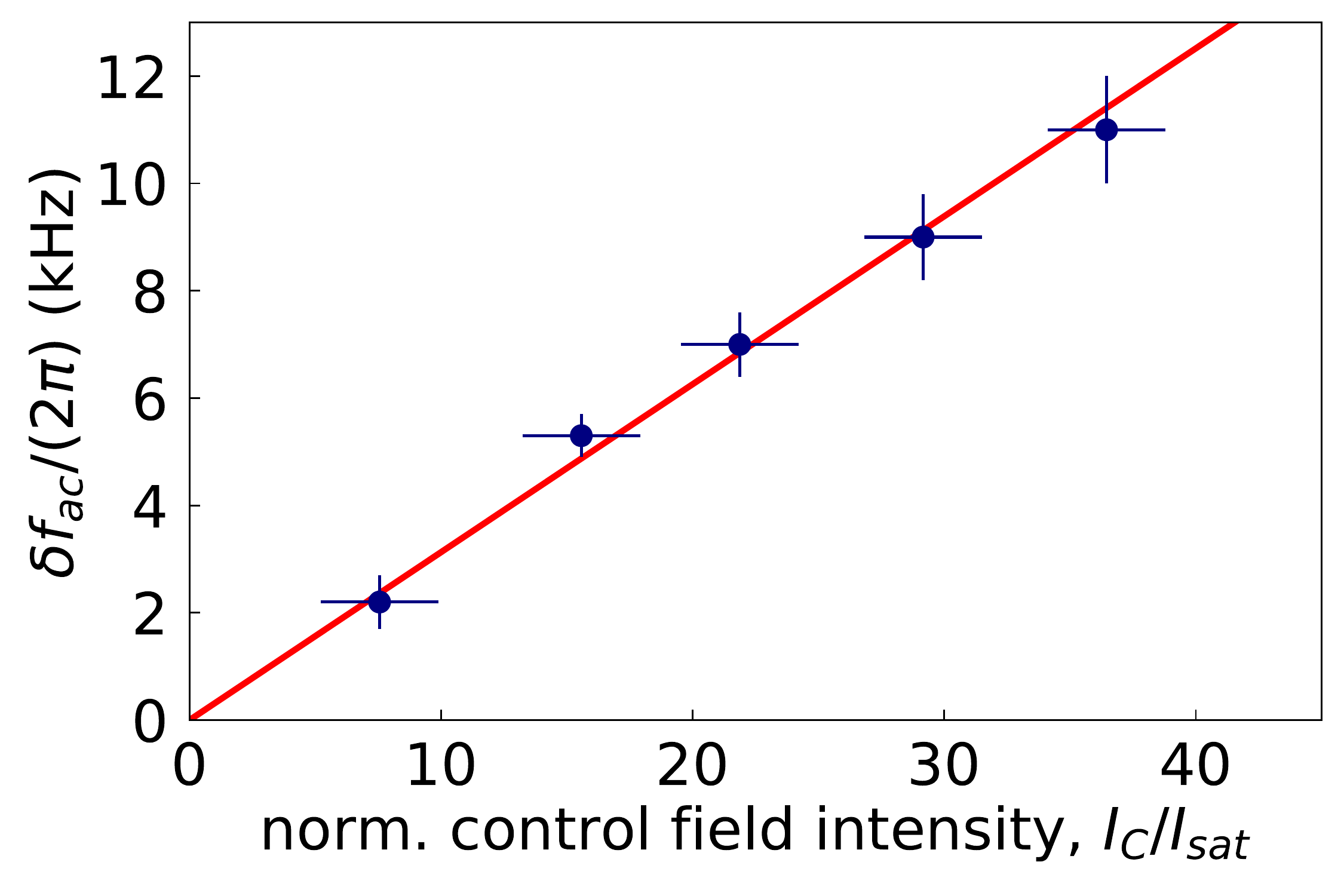}
\caption{
Light shifts measured using frequency matching in light storage experiments carried out for different intensities $ I_C $ of the control field (here normalized to the saturation intensity $ I_{sat} $). The observed ac Stark shift $ \delta f_{ac} $ (blue data points) shows good agreement with a linear dependence on $ I_C $ as indicated by a fit with a linear function (red line). The depicted uncertainties are attributed to fluctuations of $ I_C $ and standard errors of the beating frequencies calculated by the fitting routine for individual values of $ I_C $.
 }
\label{fig:IcShift}
\end{figure}

In initial experiments, we investigate dark resonances and the storage of light. Fig. \ref{fig:schematic}(a) shows the transmission of the signal beam as a function of the two-photon detuning $ 2 \pi \delta_R = \omega_S -  \omega_C - 2 g_F\mu_B B / \hbar$ at a fixed magnetic bias field, here chosen to be $ B_0 \approx 0.49 ~\text{G} $. $ \omega_S,  \omega_C, g_F, \mu_B $ and $\hbar $ denote the angular frequencies of the optical signal and control fields, the Land\'{e} $ g $ factor, the Bohr magneton, and the reduced Planck constant respectively. With typical beam powers of the control and the signal beam of $P_C = 300 ~ \mu \text{W} $ and $P_S = 100 ~ \mu \text{W} $, respectively, we observe a transmission window of approximately $ 20 ~ \text{kHz} $. 

As illustrated in Fig. \ref{fig:schematic}(b), the rubidium ensemble is irradiated with a sequence of control and signal field pulses. 
Firstly, the atoms are exposed to the control field only in order to prepare the population in the state $\ket{g_-}$ by means of optical pumping. Subsequently, a signal field field pulse with a typical duration of approximately $1~\text{ms}$ ($50~ \mu \text{s} $ for the case shown in Fig. \ref{fig:schematic}) is sent into the prepared medium, with its falling edge coinciding with that of the control field. After a storage period of $ \tau = 5~ \mu \text{s} $, the control field is turned on again to initiate the retrieval of the signal field. Fig. \ref{fig:schematic}(c) shows a typical example of such an experiment with the rapid oscillations seen both in the input and the retrieved parts of the signal being due to the optical beating with the control beam. To this end, the quarter-wave plate behind the rubidium apparatus is slightly rotated to allow leakage of the control beam onto the photodiode. The oscillation frequency $ f_B $ of the beating signal is then determined by fitting a sinusoidally modulated function to the recorded data \cite{Karpa2009}.

Under EIT conditions, at a fixed magnetic bias field, the storage experiment is repeated for different values of $ \delta_R $ within the transmission window shown in Fig. \ref{fig:schematic}(a). Fig. \ref{fig:schematic}(d) shows the extracted beat frequencies $ f_B $ of the control and input as well as retrieved signal fields, as a function of $ \delta_R $. While for the input pulse, $ f_B $ is determined by the difference between the control and signal fields' optical frequencies, the beat frequency of the retrieved signal remains constant within our experimental uncertainties, in agreement with the findings reported in \cite{Karpa2008,Karpa2009,Djokic2015}. This locking to the atomic resonance within the EIT window therefore acts as a spectroscopic tool for determining the two-photon resonance frequency even if the resonance condition is violated, that is for $ \delta_R \neq 0 $. We use the observed shift $ \delta f_{ac} $ of the intersection point away from $ \delta_R = 0 $ (illustrated by the vertical shaded area) as a measure for the light shift caused by the control field. We note that since the observed beating originates directly from the atomic coherence probed exclusively during the retrieval phase, the spectral position of the Raman resonance does not directly depend on either the width or the shape of the EIT resonance, highlighting a major advantage of our measurement scheme. Dark resonances are in general asymmetric as shown e.g. in Fig. \ref{fig:schematic}(a) and can exhibit other significant deviations e.g. from Lorentzian profiles in a non-trivial fashion which have to be taken into account when measuring the splitting between the ground states by means of recording dark resonances in the typical regime of continuous exposure to optical fields and subsequently extracting their center positions.

We now investigate the dependence of the observed shift on the intensity of the control field $ I_C $. To this end, we repeat the series of light storage experiments as depicted in Fig. \ref{fig:schematic}(d) for different values of $ I_C $ while keeping the magnetic bias field and the signal beam intensities constant. Fig. \ref{fig:IcShift} shows the result of such measurements for a signal beam power of 100 $\mu$W. We find that the ac Stark shifts observed in our experiment follow the expected linear dependence on $ I_C $ \cite{Arditi1961,Dalibard1989,Miletic2012,Metcalf1999}, which we confirm by fitting a linear function to our data. The obtained results are in good agreement with the light shifts calculated assuming our simplified multi-level system and additional coupling of the control field to the $5^2P_{1/2}, ~ F=2$ state. However, predicting the exact shift for our thermal ensembles requires more precise knowledge of numerous relevant experimental parameters \cite{Arditi1961,Wynands1999,Vanier2005}. For example, the temperature at the location of the atoms, the pressure shift, inhomogeneity of the magnetic field, impurities of the polarizations, alignment of k-vectors with the magnetic field, and the exact distribution of population between the Zeeman sublevels are known to have a direct impact on the expected light shift and have to be taken into account \cite{Wynands1999,Steck2003,Pollock2018}. We note that in the current implementation of our scheme, no measures are taken to ensure optimal pulse shaping for light storage or retrieval \cite{Novikova2007}. In particular, due to the rectangular shape of the pulses used in our experiments, the adiabaticity condition is not necessarily fulfilled such that the population distribution in the Zeeman manifold of the $ 5S_{1/2}, ~ F=2 $ state can significantly differ from the equilibrium distribution established under continuous EIT conditions. %

Notwithstanding, our method provides a means for measuring the \textit{effective differential} shift between the two ground states experienced by the atoms, including all experimental imperfections, which can be instrumental in characterizing the interaction of atomic ensembles with optical fields. As an example, in the general case of non-collinear signal and control fields intersecting at an angle $ \alpha $ the frequency pulling is determined by the relation $ \delta(\alpha) = \delta_R (1 - \cos(\alpha)) \cos^2(\Theta) $ \cite{Karpa2009}, where $ \Theta $ is the mixing angle given by $ \tan(\Theta) = g \sqrt{N} / \Omega_C $, with 
$ g $ being the coupling strength of the signal field, $ N $ the atomic density, and $ \Omega_C $ the Rabi frequency of the control field. Therefore, a comparison of the obtained shift with a theoretically predicted value or the collinear configuration, i.e. $ \alpha = 0 $, can be used to determine $ \alpha $ at the location of the atomic ensemble.

Another advantage of using light storage spectroscopy for measuring light shifts is the expected insensitivity of the beat frequency with respect to the intensity of the signal field as well as its fluctuations. This can be understood intuitively in the dark state-polariton picture, where the propagation of signal field pulses is described by the quantum field \cite{Fleischhauer2000} $ \hat{\Psi} = \cos(\Theta) \hat{E}_S + \sin(\Theta) \sqrt{N} \hat{\sigma}_{-+}^{j}$. Here, $ \hat{E}_S $ denotes the creation operator for a photon in the signal field and $ \hat{\sigma}_{-+}^{j} $ the spin operator representing the change of the state of the $ j $th atom from $ \ket{g_-} $ to $ \ket{g_+} $, respectively. As a fundamental consequence of the light storage process, no external signal fields are present during the storage period ($ \Theta = \pi /2 $), and all subsequently retrieved photons in the signal beam mode are created by the interaction of the control (read-out) field with the prepared atomic spin-wave component of the polariton.

We now test this prediction by performing light storage experiments as discussed in the context of Fig. \ref{fig:IcShift}, but for different signal field intensities while the control field power is kept constant at $ P_C \approx 300~\mu\text{W} $.
\begin{figure}[h!!!]
	\centering
	\includegraphics[width = 0.45 \textwidth]{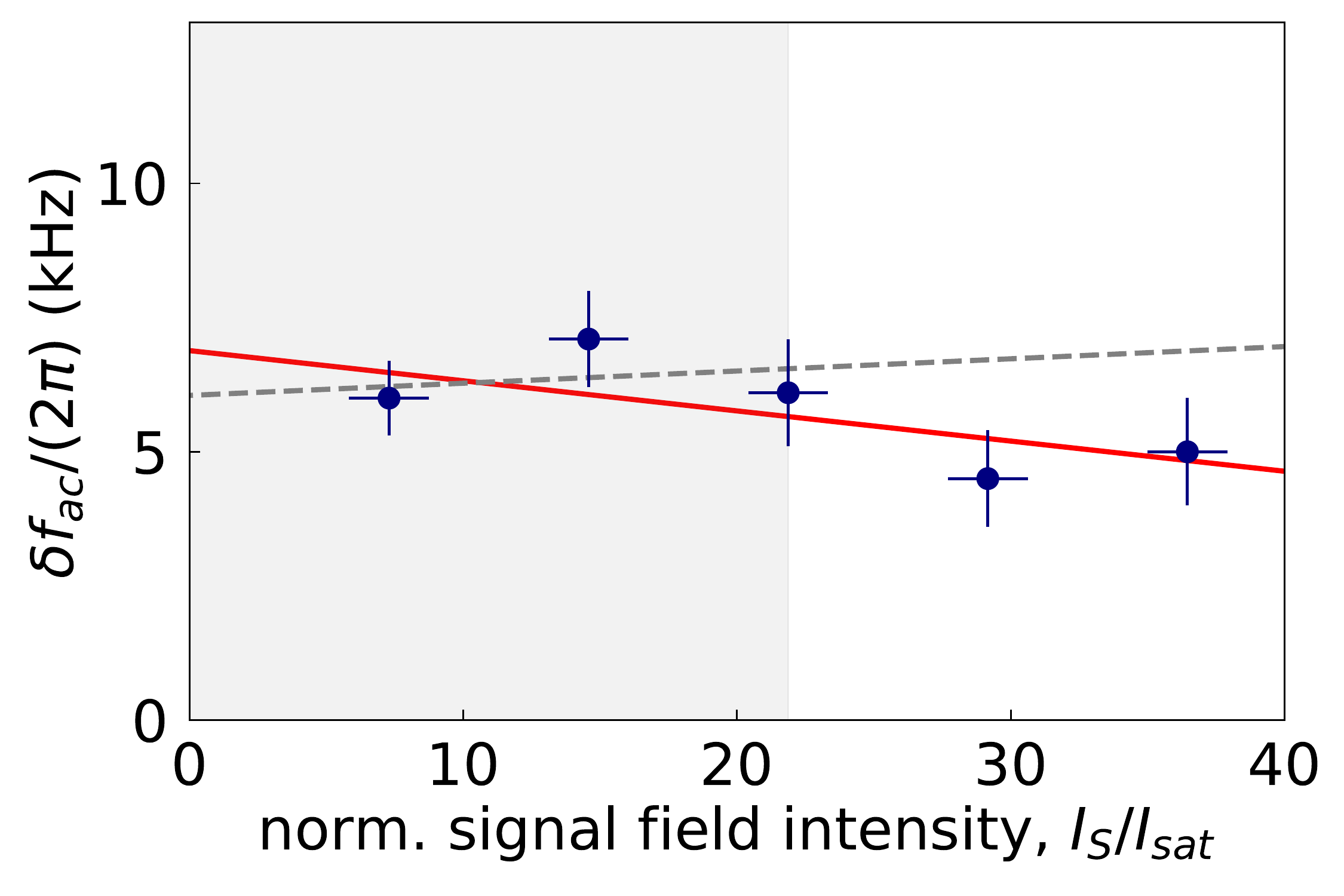}
	\caption{
		Light shift as a function of the normalized input signal field intensity. For a fixed magnetic bias field and control field intensity, the storage experiment is performed for different $ I_S $. A linear fit to all extracted frequency shifts (red solid line) and to the data within the grey shaded area indicating the range $ I_S \leq I_C $ (grey dashed line), yields a slope comparable to the corresponding standard error and a slope that is consistent with zero, respectively.
	}
	\label{fig:IsShift}
\end{figure}
Fig. \ref{fig:IsShift} shows such beat frequency measurements carried out again at a fixed magnetic bias field. In the case of signal field intensities comparable to values used in the previous measurement, we obtain $ \delta f_{ac}/(2 \pi) \approx 7 ~ \text{kHz}$, in agreement with the result shown in Fig. \ref{fig:IcShift}. However, for varied $ I_S $, we now observe that $ f_B $ remains constant within our experimental uncertainty. That is, a linear fit to the data yields a slope comparable to the corresponding standard error and approximately 30 times smaller than the value obtained for the case of varied control field intensity after accounting for the different Clebsch-Gordan coefficients of the respective optical transitions. Moreover, for signal field intensities below the point where the approximation of weak signal fields breaks down, i.e. $ I_S \approx I_C $, as indicated by the shaded area, our analysis reveals a slope consistent with zero. In the regime where $ I_S $ exceeds $ I_C $, a residual non-vanishing slope could be the result of non-trivial population dynamics in the ground state manifold occurring during the preparation phase, leading to a distribution encountered upon readout that is no longer confined to the magnetic quantum numbers $ m_F = -2 $ and $ m_F = 0 $ as assumed for the idealized $ \Lambda $ system. 
In addition, due to strongly enhanced four-wave mixing observed in EIT media \cite{Kash1999,Camacho2009}, additional optical fields can be generated during the propagation of the signal field pulses, which in turn would manifest themselves as sidebands in the beating signal. Since our current analysis only takes into account a single spectral component of the beatnote, the presence of such sidebands can lead to systematic deviations that become more pronounced with increasing signal field intensity.   
Nonetheless, this finding is in clear contrast to the demonstrated statistically significant linear dependence of the light shift with respect to the intensity of the control field, confirming that the energetic shift in our light-storage scheme is predominantly or solely caused by the control field. Consequently, extending light-storage spectroscopy to tripod-type systems comprising four electronic levels coupled by two signal fields and one control field may allow to measure the energetic splitting between the outer electronic ground states \cite{Karpa2008} while effectively avoiding the impact of light shifts. This is a consequence of the signal being derived from the beat note between the two signal fields which, upon retrieval initiated by a common control field, are shifted by the same amount. Along similar lines, if the signal fields are chosen to couple magnetically sensitive ground states, e.g. different Zeeman sublevels within a hyperfine manifold, the same methods could be applied to measure either magnetic fields \cite{Karpa2008} or magnetic field gradients \cite{Karpa2010} with improved precision.

To summarize, we have shown that light-storage spectroscopy can be used to measure control field-induced differential light shifts between electronic ground states while at he same time measuring their energetic splitting. 
The method is readily applicable to all conventional CPT-based light shift measurement schemes using continuous exposure to signal and control fields, even for strongly asymmetric transmission profiles, since it only requires minor modifications regarding the capability to operate the optical fields in a pulsed fashion.
Lastly, we show that the light shift is independent of the intensity of the signal field within our experimental uncertainties, potentially allowing to improve the precision of clocks, magnetometers or magnetogradiometers exploiting CPT or EIT in suitable three-level systems. We also discuss how the influence of ac Stark shifts in such precision measurements might be suppressed by using previously demonstrated extensions to tripod-type four-level configurations. It will be interesting to test these predictions in future experiments featuring a more refined control of the system, ideally carried out with ultracold atomic ensembles allowing for drastically improved magnetic field homogeneity and eliminating buffer gas shifts, and where many relevant parameters can be determined with very high precision. For instance, ion Coulomb crystals are a system holding great promise for improving the coherence time and light storage efficiency required for precision measurements \cite{Albert2011} owing to favourable properties such as intrinsically suppressed collisional decoherence, while potential detrimental effects stemming from the presence of micromotion \cite{Cirac1994,Ludlow2015a} might be avoided by using optical trapping techniques \cite{Lambrecht2017,Schaetz2017,Schmidt2018,Karpa2019,Schmidt2020}. 
%

\begin{thebibliography}{36}%
	\makeatletter
	\providecommand \@ifxundefined [1]{%
		\@ifx{#1\undefined}
	}%
	\providecommand \@ifnum [1]{%
		\ifnum #1\expandafter \@firstoftwo
		\else \expandafter \@secondoftwo
		\fi
	}%
	\providecommand \@ifx [1]{%
		\ifx #1\expandafter \@firstoftwo
		\else \expandafter \@secondoftwo
		\fi
	}%
	\providecommand \natexlab [1]{#1}%
	\providecommand \enquote  [1]{``#1''}%
	\providecommand \bibnamefont  [1]{#1}%
	\providecommand \bibfnamefont [1]{#1}%
	\providecommand \citenamefont [1]{#1}%
	\providecommand \href@noop [0]{\@secondoftwo}%
	\providecommand \href [0]{\begingroup \@sanitize@url \@href}%
	\providecommand \@href[1]{\@@startlink{#1}\@@href}%
	\providecommand \@@href[1]{\endgroup#1\@@endlink}%
	\providecommand \@sanitize@url [0]{\catcode `\\12\catcode `\$12\catcode
		`\&12\catcode `\#12\catcode `\^12\catcode `\_12\catcode `\%12\relax}%
	\providecommand \@@startlink[1]{}%
	\providecommand \@@endlink[0]{}%
	\providecommand \url  [0]{\begingroup\@sanitize@url \@url }%
	\providecommand \@url [1]{\endgroup\@href {#1}{\urlprefix }}%
	\providecommand \urlprefix  [0]{URL }%
	\providecommand \Eprint [0]{\href }%
	\providecommand \doibase [0]{https://doi.org/}%
	\providecommand \selectlanguage [0]{\@gobble}%
	\providecommand \bibinfo  [0]{\@secondoftwo}%
	\providecommand \bibfield  [0]{\@secondoftwo}%
	\providecommand \translation [1]{[#1]}%
	\providecommand \BibitemOpen [0]{}%
	\providecommand \bibitemStop [0]{}%
	\providecommand \bibitemNoStop [0]{.\EOS\space}%
	\providecommand \EOS [0]{\spacefactor3000\relax}%
	\providecommand \BibitemShut  [1]{\csname bibitem#1\endcsname}%
	\let\auto@bib@innerbib\@empty
	\bibitem [{\citenamefont {Alzetta}\ \emph {et~al.}(1976)\citenamefont
		{Alzetta}, \citenamefont {Gozzini}, \citenamefont {Moi},\ and\ \citenamefont
		{Orriols}}]{Alzetta1976}%
	\BibitemOpen
	\bibfield  {author} {\bibinfo {author} {\bibfnamefont {G.}~\bibnamefont
			{Alzetta}}, \bibinfo {author} {\bibfnamefont {A.}~\bibnamefont {Gozzini}},
		\bibinfo {author} {\bibfnamefont {L.}~\bibnamefont {Moi}},\ and\ \bibinfo
		{author} {\bibfnamefont {G.}~\bibnamefont {Orriols}},\ }\bibfield  {title}
	{\bibinfo {title} {{An experimental method for the observation of r.f.
				transitions and laser beat resonances in oriented Na vapour}},\ }\href
	{https://doi.org/10.1007/BF02749417} {\bibfield  {journal} {\bibinfo
			{journal} {Il Nuovo Cimento B Series 11}\ }\textbf {\bibinfo {volume} {36}},\
		\bibinfo {pages} {5} (\bibinfo {year} {1976})}\BibitemShut {NoStop}%
	\bibitem [{\citenamefont {Arimondo}(1996)}]{Arimondo1996}%
	\BibitemOpen
	\bibfield  {author} {\bibinfo {author} {\bibfnamefont {E.}~\bibnamefont
			{Arimondo}},\ }\bibfield  {title} {\bibinfo {title} {Coherent population
			trapping in laser spectroscopy}\ }(\bibinfo  {publisher} {Elsevier},\
	\bibinfo {year} {1996})\ pp.\ \bibinfo {pages} {257 -- 354}\BibitemShut
	{NoStop}%
	\bibitem [{\citenamefont {Harris}\ \emph {et~al.}(1990)\citenamefont {Harris},
		\citenamefont {Field},\ and\ \citenamefont {Imamo\ifmmode~\breve{g}\else
			\u{g}\fi{}lu}}]{Harris1990}%
	\BibitemOpen
	\bibfield  {author} {\bibinfo {author} {\bibfnamefont {S.~E.}\ \bibnamefont
			{Harris}}, \bibinfo {author} {\bibfnamefont {J.~E.}\ \bibnamefont {Field}},\
		and\ \bibinfo {author} {\bibfnamefont {A.}~\bibnamefont
			{Imamo\ifmmode~\breve{g}\else \u{g}\fi{}lu}},\ }\bibfield  {title} {\bibinfo
		{title} {Nonlinear optical processes using electromagnetically induced
			transparency},\ }\href {https://doi.org/10.1103/PhysRevLett.64.1107}
	{\bibfield  {journal} {\bibinfo  {journal} {Phys. Rev. Lett.}\ }\textbf
		{\bibinfo {volume} {64}},\ \bibinfo {pages} {1107} (\bibinfo {year}
		{1990})}\BibitemShut {NoStop}%
	\bibitem [{\citenamefont {Fleischhauer}\ \emph {et~al.}(2005)\citenamefont
		{Fleischhauer}, \citenamefont {Imamoglu},\ and\ \citenamefont
		{Marangos}}]{Fleischhauer2005}%
	\BibitemOpen
	\bibfield  {author} {\bibinfo {author} {\bibfnamefont {M.}~\bibnamefont
			{Fleischhauer}}, \bibinfo {author} {\bibfnamefont {A.}~\bibnamefont
			{Imamoglu}},\ and\ \bibinfo {author} {\bibfnamefont {J.~P.}\ \bibnamefont
			{Marangos}},\ }\bibfield  {title} {\bibinfo {title} {{Electromagnetically
				induced transparency: Optics in coherent media}},\ }\href
	{https://doi.org/10.1103/RevModPhys.77.633} {\bibfield  {journal} {\bibinfo
			{journal} {Reviews of Modern Physics}\ }\textbf {\bibinfo {volume} {77}},\
		\bibinfo {pages} {633} (\bibinfo {year} {2005})},\ \Eprint
	{https://arxiv.org/abs/0001094} {arXiv:0001094 [quant-ph]} \BibitemShut
	{NoStop}%
	\bibitem [{\citenamefont {Brandt}\ \emph {et~al.}(1997)\citenamefont {Brandt},
		\citenamefont {Nagel}, \citenamefont {Wynands},\ and\ \citenamefont
		{Meschede}}]{Brandt1997}%
	\BibitemOpen
	\bibfield  {author} {\bibinfo {author} {\bibfnamefont {S.}~\bibnamefont
			{Brandt}}, \bibinfo {author} {\bibfnamefont {A.}~\bibnamefont {Nagel}},
		\bibinfo {author} {\bibfnamefont {R.}~\bibnamefont {Wynands}},\ and\ \bibinfo
		{author} {\bibfnamefont {D.}~\bibnamefont {Meschede}},\ }\bibfield  {title}
	{\bibinfo {title} {{Buffer-gas-induced linewidth reduction of coherent dark
				resonances to below $50 $ Hz}},\ }\href
	{https://doi.org/10.1103/PhysRevA.56.R1063} {\bibfield  {journal} {\bibinfo
			{journal} {Phys. Rev. A}\ }\textbf {\bibinfo {volume} {56}},\ \bibinfo
		{pages} {R1063} (\bibinfo {year} {1997})}\BibitemShut {NoStop}%
	\bibitem [{\citenamefont {Hau}\ \emph {et~al.}(1999)\citenamefont {Hau},
		\citenamefont {Harris}, \citenamefont {Dutton},\ and\ \citenamefont
		{Behroozi}}]{Hau1999}%
	\BibitemOpen
	\bibfield  {author} {\bibinfo {author} {\bibfnamefont {L.~V.}\ \bibnamefont
			{Hau}}, \bibinfo {author} {\bibfnamefont {S.~E.}\ \bibnamefont {Harris}},
		\bibinfo {author} {\bibfnamefont {Z.}~\bibnamefont {Dutton}},\ and\ \bibinfo
		{author} {\bibfnamefont {C.~H.}\ \bibnamefont {Behroozi}},\ }\bibfield
	{title} {\bibinfo {title} {{Light speed reduction to 17 metres per second in
				an ultracold atomic gas}},\ }\href {https://doi.org/10.1038/17561} {\bibfield
		{journal} {\bibinfo  {journal} {Nature}\ }\textbf {\bibinfo {volume}
			{397}},\ \bibinfo {pages} {594} (\bibinfo {year} {1999})}\BibitemShut
	{NoStop}%
	\bibitem [{\citenamefont {Kash}\ \emph {et~al.}(1999)\citenamefont {Kash},
		\citenamefont {Sautenkov}, \citenamefont {Zibrov}, \citenamefont {Hollberg},
		\citenamefont {Welch}, \citenamefont {Lukin}, \citenamefont {Rostovtsev},
		\citenamefont {Fry},\ and\ \citenamefont {Scully}}]{Kash1999}%
	\BibitemOpen
	\bibfield  {author} {\bibinfo {author} {\bibfnamefont {M.}~\bibnamefont
			{Kash}}, \bibinfo {author} {\bibfnamefont {V.}~\bibnamefont {Sautenkov}},
		\bibinfo {author} {\bibfnamefont {A.}~\bibnamefont {Zibrov}}, \bibinfo
		{author} {\bibfnamefont {L.}~\bibnamefont {Hollberg}}, \bibinfo {author}
		{\bibfnamefont {G.}~\bibnamefont {Welch}}, \bibinfo {author} {\bibfnamefont
			{M.}~\bibnamefont {Lukin}}, \bibinfo {author} {\bibfnamefont
			{Y.}~\bibnamefont {Rostovtsev}}, \bibinfo {author} {\bibfnamefont
			{E.}~\bibnamefont {Fry}},\ and\ \bibinfo {author} {\bibfnamefont
			{M.}~\bibnamefont {Scully}},\ }\bibfield  {title} {\bibinfo {title}
		{{Ultraslow Group Velocity and Enhanced Nonlinear Optical Effects in a
				Coherently Driven Hot Atomic Gas}},\ }\href
	{https://doi.org/10.1103/PhysRevLett.82.5229} {\bibfield  {journal} {\bibinfo
			{journal} {Physical Review Letters}\ }\textbf {\bibinfo {volume} {82}},\
		\bibinfo {pages} {5229} (\bibinfo {year} {1999})}\BibitemShut {NoStop}%
	\bibitem [{\citenamefont {Budker}\ \emph {et~al.}(1999)\citenamefont {Budker},
		\citenamefont {Kimball}, \citenamefont {Rochester},\ and\ \citenamefont
		{Yashchuk}}]{Budker1999}%
	\BibitemOpen
	\bibfield  {author} {\bibinfo {author} {\bibfnamefont {D.}~\bibnamefont
			{Budker}}, \bibinfo {author} {\bibfnamefont {D.~F.}\ \bibnamefont {Kimball}},
		\bibinfo {author} {\bibfnamefont {S.~M.}\ \bibnamefont {Rochester}},\ and\
		\bibinfo {author} {\bibfnamefont {V.~V.}\ \bibnamefont {Yashchuk}},\
	}\bibfield  {title} {\bibinfo {title} {{Nonlinear Magneto-optics and Reduced
				Group Velocity of Light in Atomic Vapor with Slow Ground State Relaxation}},\
	}\href {https://doi.org/10.1103/PhysRevLett.83.1767} {\bibfield  {journal}
		{\bibinfo  {journal} {Physical Review Letters}\ }\textbf {\bibinfo {volume}
			{83}},\ \bibinfo {pages} {1767} (\bibinfo {year} {1999})}\BibitemShut
	{NoStop}%
	\bibitem [{\citenamefont {Scully}\ and\ \citenamefont
		{Fleischhauer}(1992)}]{Scully1992}%
	\BibitemOpen
	\bibfield  {author} {\bibinfo {author} {\bibfnamefont {M.~O.}\ \bibnamefont
			{Scully}}\ and\ \bibinfo {author} {\bibfnamefont {M.}~\bibnamefont
			{Fleischhauer}},\ }\bibfield  {title} {\bibinfo {title} {High-sensitivity
			magnetometer based on index-enhanced media},\ }\href
	{https://doi.org/10.1103/PhysRevLett.69.1360} {\bibfield  {journal} {\bibinfo
			{journal} {Phys. Rev. Lett.}\ }\textbf {\bibinfo {volume} {69}},\ \bibinfo
		{pages} {1360} (\bibinfo {year} {1992})}\BibitemShut {NoStop}%
	\bibitem [{\citenamefont {Katsoprinakis}\ \emph {et~al.}(2006)\citenamefont
		{Katsoprinakis}, \citenamefont {Petrosyan},\ and\ \citenamefont
		{Kominis}}]{Katsoprinakis2006}%
	\BibitemOpen
	\bibfield  {author} {\bibinfo {author} {\bibfnamefont {G.}~\bibnamefont
			{Katsoprinakis}}, \bibinfo {author} {\bibfnamefont {D.}~\bibnamefont
			{Petrosyan}},\ and\ \bibinfo {author} {\bibfnamefont {I.~K.}\ \bibnamefont
			{Kominis}},\ }\bibfield  {title} {\bibinfo {title} {{High Frequency Atomic
				Magnetometer by Use of Electromagnetically Induced Transparency}},\ }\href
	{https://doi.org/10.1103/PhysRevLett.97.230801} {\bibfield  {journal}
		{\bibinfo  {journal} {Phys. Rev. Lett.}\ }\textbf {\bibinfo {volume} {97}},\
		\bibinfo {pages} {230801} (\bibinfo {year} {2006})}\BibitemShut {NoStop}%
	\bibitem [{\citenamefont {Vanier}(2005)}]{Vanier2005}%
	\BibitemOpen
	\bibfield  {author} {\bibinfo {author} {\bibfnamefont {J.}~\bibnamefont
			{Vanier}},\ }\bibfield  {title} {\bibinfo {title} {{Atomic clocks based on
				coherent population trapping: a review}},\ }\href
	{https://doi.org/10.1007/s00340-005-1905-3} {\bibfield  {journal} {\bibinfo
			{journal} {Applied Physics B}\ }\textbf {\bibinfo {volume} {81}},\ \bibinfo
		{pages} {421} (\bibinfo {year} {2005})}\BibitemShut {NoStop}%
	\bibitem [{\citenamefont {Ludlow}\ \emph {et~al.}(2015)\citenamefont {Ludlow},
		\citenamefont {Boyd}, \citenamefont {Ye}, \citenamefont {Peik},\ and\
		\citenamefont {Schmidt}}]{Ludlow2015a}%
	\BibitemOpen
	\bibfield  {author} {\bibinfo {author} {\bibfnamefont {A.~D.}\ \bibnamefont
			{Ludlow}}, \bibinfo {author} {\bibfnamefont {M.~M.}\ \bibnamefont {Boyd}},
		\bibinfo {author} {\bibfnamefont {J.}~\bibnamefont {Ye}}, \bibinfo {author}
		{\bibfnamefont {E.}~\bibnamefont {Peik}},\ and\ \bibinfo {author}
		{\bibfnamefont {P.~O.}\ \bibnamefont {Schmidt}},\ }\bibfield  {title}
	{\bibinfo {title} {Optical atomic clocks},\ }\href
	{https://doi.org/10.1103/RevModPhys.87.637} {\bibfield  {journal} {\bibinfo
			{journal} {Rev. Mod. Phys.}\ }\textbf {\bibinfo {volume} {87}},\ \bibinfo
		{pages} {637} (\bibinfo {year} {2015})}\BibitemShut {NoStop}%
	\bibitem [{\citenamefont {Zanon-Willette}\ \emph {et~al.}(2018)\citenamefont
		{Zanon-Willette}, \citenamefont {Lefevre}, \citenamefont {Metzdorff},
		\citenamefont {Sillitoe}, \citenamefont {Almonacil}, \citenamefont
		{Minissale}, \citenamefont {de~Clercq}, \citenamefont {Taichenachev},
		\citenamefont {Yudin},\ and\ \citenamefont {Arimondo}}]{ZanonWillette2018}%
	\BibitemOpen
	\bibfield  {author} {\bibinfo {author} {\bibfnamefont {T.}~\bibnamefont
			{Zanon-Willette}}, \bibinfo {author} {\bibfnamefont {R.}~\bibnamefont
			{Lefevre}}, \bibinfo {author} {\bibfnamefont {R.}~\bibnamefont {Metzdorff}},
		\bibinfo {author} {\bibfnamefont {N.}~\bibnamefont {Sillitoe}}, \bibinfo
		{author} {\bibfnamefont {S.}~\bibnamefont {Almonacil}}, \bibinfo {author}
		{\bibfnamefont {M.}~\bibnamefont {Minissale}}, \bibinfo {author}
		{\bibfnamefont {E.}~\bibnamefont {de~Clercq}}, \bibinfo {author}
		{\bibfnamefont {A.~V.}\ \bibnamefont {Taichenachev}}, \bibinfo {author}
		{\bibfnamefont {V.~I.}\ \bibnamefont {Yudin}},\ and\ \bibinfo {author}
		{\bibfnamefont {E.}~\bibnamefont {Arimondo}},\ }\bibfield  {title} {\bibinfo
		{title} {Composite laser-pulses spectroscopy for high-accuracy optical
			clocks: a review of recent progress and perspectives},\ }\href
	{https://doi.org/10.1088/1361-6633/aac9e9} {\bibfield  {journal} {\bibinfo
			{journal} {Reports on Progress in Physics}\ }\textbf {\bibinfo {volume}
			{81}},\ \bibinfo {pages} {094401} (\bibinfo {year} {2018})}\BibitemShut
	{NoStop}%
	\bibitem [{\citenamefont {Fleischhauer}\ and\ \citenamefont
		{Lukin}(2000)}]{Fleischhauer2000}%
	\BibitemOpen
	\bibfield  {author} {\bibinfo {author} {\bibfnamefont {M.}~\bibnamefont
			{Fleischhauer}}\ and\ \bibinfo {author} {\bibfnamefont {M.~D.}\ \bibnamefont
			{Lukin}},\ }\bibfield  {title} {\bibinfo {title} {{Dark-state polaritons in
				electromagnetically induced transparency}},\ }\href
	{https://doi.org/10.1103/PhysRevLett.84.5094} {\bibfield  {journal} {\bibinfo
			{journal} {Phys. Rev. Lett.}\ }\textbf {\bibinfo {volume} {84}},\ \bibinfo
		{pages} {5094} (\bibinfo {year} {2000})},\ \Eprint
	{https://arxiv.org/abs/0001094} {arXiv:0001094 [quant-ph]} \BibitemShut
	{NoStop}%
	\bibitem [{\citenamefont {Liu}\ \emph {et~al.}(2001)\citenamefont {Liu},
		\citenamefont {Dutton}, \citenamefont {Behroozi},\ and\ \citenamefont
		{Hau}}]{Liu2001}%
	\BibitemOpen
	\bibfield  {author} {\bibinfo {author} {\bibfnamefont {C.}~\bibnamefont
			{Liu}}, \bibinfo {author} {\bibfnamefont {Z.}~\bibnamefont {Dutton}},
		\bibinfo {author} {\bibfnamefont {C.~H.}\ \bibnamefont {Behroozi}},\ and\
		\bibinfo {author} {\bibfnamefont {L.~V.}\ \bibnamefont {Hau}},\ }\bibfield
	{title} {\bibinfo {title} {{Observation of coherent optical information
				storage in an atomic medium using halted light pulses}},\ }\href
	{https://doi.org/10.1038/35054017} {\bibfield  {journal} {\bibinfo  {journal}
			{Nature}\ }\textbf {\bibinfo {volume} {409}},\ \bibinfo {pages} {490}
		(\bibinfo {year} {2001})}\BibitemShut {NoStop}%
	\bibitem [{\citenamefont {Phillips}\ \emph {et~al.}(2001)\citenamefont
		{Phillips}, \citenamefont {Fleischhauer}, \citenamefont {Mair}, \citenamefont
		{Walsworth},\ and\ \citenamefont {Lukin}}]{Phillips2001}%
	\BibitemOpen
	\bibfield  {author} {\bibinfo {author} {\bibfnamefont {D.~F.}\ \bibnamefont
			{Phillips}}, \bibinfo {author} {\bibfnamefont {A.}~\bibnamefont
			{Fleischhauer}}, \bibinfo {author} {\bibfnamefont {A.}~\bibnamefont {Mair}},
		\bibinfo {author} {\bibfnamefont {R.~L.}\ \bibnamefont {Walsworth}},\ and\
		\bibinfo {author} {\bibfnamefont {M.~D.}\ \bibnamefont {Lukin}},\ }\bibfield
	{title} {\bibinfo {title} {{Storage of Light in Atomic Vapor}},\ }\href
	{https://doi.org/10.1103/PhysRevLett.86.783} {\bibfield  {journal} {\bibinfo
			{journal} {Phys. Rev. Lett.}\ }\textbf {\bibinfo {volume} {86}},\ \bibinfo
		{pages} {783} (\bibinfo {year} {2001})}\BibitemShut {NoStop}%
	\bibitem [{\citenamefont {Karpa}\ \emph {et~al.}(2009)\citenamefont {Karpa},
		\citenamefont {Nikoghosyan}, \citenamefont {Vewinger}, \citenamefont
		{Fleischhauer},\ and\ \citenamefont {Weitz}}]{Karpa2009}%
	\BibitemOpen
	\bibfield  {author} {\bibinfo {author} {\bibfnamefont {L.}~\bibnamefont
			{Karpa}}, \bibinfo {author} {\bibfnamefont {G.}~\bibnamefont {Nikoghosyan}},
		\bibinfo {author} {\bibfnamefont {F.}~\bibnamefont {Vewinger}}, \bibinfo
		{author} {\bibfnamefont {M.}~\bibnamefont {Fleischhauer}},\ and\ \bibinfo
		{author} {\bibfnamefont {M.}~\bibnamefont {Weitz}},\ }\bibfield  {title}
	{\bibinfo {title} {Frequency matching in light-storage spectroscopy of atomic
			raman transitions},\ }\href {https://doi.org/10.1103/PhysRevLett.103.093601}
	{\bibfield  {journal} {\bibinfo  {journal} {Phys. Rev. Lett.}\ }\textbf
		{\bibinfo {volume} {103}},\ \bibinfo {pages} {093601} (\bibinfo {year}
		{2009})}\BibitemShut {NoStop}%
	\bibitem [{\citenamefont {Karpa}\ \emph {et~al.}(2008)\citenamefont {Karpa},
		\citenamefont {Vewinger},\ and\ \citenamefont {Weitz}}]{Karpa2008}%
	\BibitemOpen
	\bibfield  {author} {\bibinfo {author} {\bibfnamefont {L.}~\bibnamefont
			{Karpa}}, \bibinfo {author} {\bibfnamefont {F.}~\bibnamefont {Vewinger}},\
		and\ \bibinfo {author} {\bibfnamefont {M.}~\bibnamefont {Weitz}},\ }\bibfield
	{title} {\bibinfo {title} {{Resonance beating of light stored using atomic
				spinor polaritons}},\ }\href {https://doi.org/10.1103/PhysRevLett.101.170406}
	{\bibfield  {journal} {\bibinfo  {journal} {Physical Review Letters}\
		}\textbf {\bibinfo {volume} {101}},\ \bibinfo {pages} {2} (\bibinfo {year}
		{2008})},\ \Eprint {https://arxiv.org/abs/0807.3853} {arXiv:0807.3853}
	\BibitemShut {NoStop}%
	\bibitem [{\citenamefont {Karpa}\ and\ \citenamefont
		{Weitz}(2010)}]{Karpa2010}%
	\BibitemOpen
	\bibfield  {author} {\bibinfo {author} {\bibfnamefont {L.}~\bibnamefont
			{Karpa}}\ and\ \bibinfo {author} {\bibfnamefont {M.}~\bibnamefont {Weitz}},\
	}\bibfield  {title} {\bibinfo {title} {Nondispersive optics using storage of
			light},\ }\href {https://doi.org/10.1103/PhysRevA.81.041802} {\bibfield
		{journal} {\bibinfo  {journal} {Phys. Rev. A}\ }\textbf {\bibinfo {volume}
			{81}},\ \bibinfo {pages} {041802} (\bibinfo {year} {2010})}\BibitemShut
	{NoStop}%
	\bibitem [{\citenamefont {Djokic}\ \emph {et~al.}(2015)\citenamefont {Djokic},
		\citenamefont {Enzian}, \citenamefont {Vewinger},\ and\ \citenamefont
		{Weitz}}]{Djokic2015}%
	\BibitemOpen
	\bibfield  {author} {\bibinfo {author} {\bibfnamefont {V.}~\bibnamefont
			{Djokic}}, \bibinfo {author} {\bibfnamefont {G.}~\bibnamefont {Enzian}},
		\bibinfo {author} {\bibfnamefont {F.}~\bibnamefont {Vewinger}},\ and\
		\bibinfo {author} {\bibfnamefont {M.}~\bibnamefont {Weitz}},\ }\bibfield
	{title} {\bibinfo {title} {Resonance retrieval of stored coherence in an
			rf-optical double-resonance experiment},\ }\href
	{https://doi.org/10.1103/PhysRevA.92.063802} {\bibfield  {journal} {\bibinfo
			{journal} {Phys. Rev. A}\ }\textbf {\bibinfo {volume} {92}},\ \bibinfo
		{pages} {063802} (\bibinfo {year} {2015})}\BibitemShut {NoStop}%
	\bibitem [{\citenamefont {Arditi}\ and\ \citenamefont
		{Carver}(1961)}]{Arditi1961}%
	\BibitemOpen
	\bibfield  {author} {\bibinfo {author} {\bibfnamefont {M.}~\bibnamefont
			{Arditi}}\ and\ \bibinfo {author} {\bibfnamefont {T.~R.}\ \bibnamefont
			{Carver}},\ }\bibfield  {title} {\bibinfo {title} {Pressure, light, and
			temperature shifts in optical detection of 0-0 hyperfine resonance of alkali
			metals},\ }\href {https://doi.org/10.1103/PhysRev.124.800} {\bibfield
		{journal} {\bibinfo  {journal} {Phys. Rev.}\ }\textbf {\bibinfo {volume}
			{124}},\ \bibinfo {pages} {800} (\bibinfo {year} {1961})}\BibitemShut
	{NoStop}%
	\bibitem [{\citenamefont {Dalibard}\ and\ \citenamefont
		{Cohen-Tannoudji}(1989)}]{Dalibard1989}%
	\BibitemOpen
	\bibfield  {author} {\bibinfo {author} {\bibfnamefont {J.}~\bibnamefont
			{Dalibard}}\ and\ \bibinfo {author} {\bibfnamefont {C.}~\bibnamefont
			{Cohen-Tannoudji}},\ }\bibfield  {title} {\bibinfo {title} {Laser cooling
			below the doppler limit by polarization gradients: simple theoretical
			models},\ }\href {https://doi.org/10.1364/JOSAB.6.002023} {\bibfield
		{journal} {\bibinfo  {journal} {J. Opt. Soc. Am. B}\ }\textbf {\bibinfo
			{volume} {6}},\ \bibinfo {pages} {2023} (\bibinfo {year} {1989})}\BibitemShut
	{NoStop}%
	\bibitem [{\citenamefont {Miletic}\ \emph {et~al.}(2012)\citenamefont
		{Miletic}, \citenamefont {Affolderbach}, \citenamefont {Hasegawa},
		\citenamefont {Boudot}, \citenamefont {Gorecki},\ and\ \citenamefont
		{Mileti}}]{Miletic2012}%
	\BibitemOpen
	\bibfield  {author} {\bibinfo {author} {\bibfnamefont {D.}~\bibnamefont
			{Miletic}}, \bibinfo {author} {\bibfnamefont {C.}~\bibnamefont
			{Affolderbach}}, \bibinfo {author} {\bibfnamefont {M.}~\bibnamefont
			{Hasegawa}}, \bibinfo {author} {\bibfnamefont {R.}~\bibnamefont {Boudot}},
		\bibinfo {author} {\bibfnamefont {C.}~\bibnamefont {Gorecki}},\ and\ \bibinfo
		{author} {\bibfnamefont {G.}~\bibnamefont {Mileti}},\ }\bibfield  {title}
	{\bibinfo {title} {{AC Stark-shift in CPT-based Cs miniature atomic
				clocks}},\ }\href {https://doi.org/10.1007/s00340-012-5121-7} {\bibfield
		{journal} {\bibinfo  {journal} {Applied Physics B}\ }\textbf {\bibinfo
			{volume} {109}},\ \bibinfo {pages} {89} (\bibinfo {year} {2012})}\BibitemShut
	{NoStop}%
	\bibitem [{\citenamefont {Metcalf}\ and\ \citenamefont {Van~der
			Straten}(1999)}]{Metcalf1999}%
	\BibitemOpen
	\bibfield  {author} {\bibinfo {author} {\bibfnamefont {H.}~\bibnamefont
			{Metcalf}}\ and\ \bibinfo {author} {\bibfnamefont {P.}~\bibnamefont {Van~der
				Straten}},\ }\href@noop {} {\emph {\bibinfo {title} {{Laser Cooling and
					Trapping}}}}\ (\bibinfo  {publisher} {Springer},\ \bibinfo {year}
	{1999})\BibitemShut {NoStop}%
	\bibitem [{\citenamefont {Wynands}\ and\ \citenamefont
		{Nagel}(1999)}]{Wynands1999}%
	\BibitemOpen
	\bibfield  {author} {\bibinfo {author} {\bibfnamefont {R.}~\bibnamefont
			{Wynands}}\ and\ \bibinfo {author} {\bibfnamefont {A.}~\bibnamefont
			{Nagel}},\ }\bibfield  {title} {\bibinfo {title} {{Precision spectroscopy
				with coherent dark states}},\ }\href {https://doi.org/10.1007/s003400050581}
	{\bibfield  {journal} {\bibinfo  {journal} {Applied Physics B: Lasers and
				Optics}\ }\textbf {\bibinfo {volume} {68}},\ \bibinfo {pages} {1} (\bibinfo
		{year} {1999})}\BibitemShut {NoStop}%
	\bibitem [{\citenamefont {Steck}(2003)}]{Steck2003}%
	\BibitemOpen
	\bibfield  {author} {\bibinfo {author} {\bibfnamefont {D.}~\bibnamefont
			{Steck}},\ }\bibfield  {title} {\bibinfo {title} {{Rubidium 87 D Line
				Data}},\ }\href@noop {} {\  (\bibinfo {year} {2003})}\BibitemShut {NoStop}%
	\bibitem [{\citenamefont {Pollock}\ \emph {et~al.}(2018)\citenamefont
		{Pollock}, \citenamefont {Yudin}, \citenamefont {Shuker}, \citenamefont
		{Basalaev}, \citenamefont {Taichenachev}, \citenamefont {Liu}, \citenamefont
		{Kitching},\ and\ \citenamefont {Donley}}]{Pollock2018}%
	\BibitemOpen
	\bibfield  {author} {\bibinfo {author} {\bibfnamefont {J.~W.}\ \bibnamefont
			{Pollock}}, \bibinfo {author} {\bibfnamefont {V.~I.}\ \bibnamefont {Yudin}},
		\bibinfo {author} {\bibfnamefont {M.}~\bibnamefont {Shuker}}, \bibinfo
		{author} {\bibfnamefont {M.~Y.}\ \bibnamefont {Basalaev}}, \bibinfo {author}
		{\bibfnamefont {A.~V.}\ \bibnamefont {Taichenachev}}, \bibinfo {author}
		{\bibfnamefont {X.}~\bibnamefont {Liu}}, \bibinfo {author} {\bibfnamefont
			{J.}~\bibnamefont {Kitching}},\ and\ \bibinfo {author} {\bibfnamefont
			{E.~A.}\ \bibnamefont {Donley}},\ }\bibfield  {title} {\bibinfo {title} {{ac
				Stark shifts of dark resonances probed with Ramsey spectroscopy}},\ }\href
	{https://doi.org/10.1103/PhysRevA.98.053424} {\bibfield  {journal} {\bibinfo
			{journal} {Phys. Rev. A}\ }\textbf {\bibinfo {volume} {98}},\ \bibinfo
		{pages} {053424} (\bibinfo {year} {2018})}\BibitemShut {NoStop}%
	\bibitem [{\citenamefont {Novikova}\ \emph {et~al.}(2007)\citenamefont
		{Novikova}, \citenamefont {Gorshkov}, \citenamefont {Phillips}, \citenamefont
		{S\o{}rensen}, \citenamefont {Lukin},\ and\ \citenamefont
		{Walsworth}}]{Novikova2007}%
	\BibitemOpen
	\bibfield  {author} {\bibinfo {author} {\bibfnamefont {I.}~\bibnamefont
			{Novikova}}, \bibinfo {author} {\bibfnamefont {A.~V.}\ \bibnamefont
			{Gorshkov}}, \bibinfo {author} {\bibfnamefont {D.~F.}\ \bibnamefont
			{Phillips}}, \bibinfo {author} {\bibfnamefont {A.~S.}\ \bibnamefont
			{S\o{}rensen}}, \bibinfo {author} {\bibfnamefont {M.~D.}\ \bibnamefont
			{Lukin}},\ and\ \bibinfo {author} {\bibfnamefont {R.~L.}\ \bibnamefont
			{Walsworth}},\ }\bibfield  {title} {\bibinfo {title} {{Optimal Control of
				Light Pulse Storage and Retrieval}},\ }\href
	{https://doi.org/10.1103/PhysRevLett.98.243602} {\bibfield  {journal}
		{\bibinfo  {journal} {Phys. Rev. Lett.}\ }\textbf {\bibinfo {volume} {98}},\
		\bibinfo {pages} {243602} (\bibinfo {year} {2007})}\BibitemShut {NoStop}%
	\bibitem [{\citenamefont {Camacho}\ \emph {et~al.}(2009)\citenamefont
		{Camacho}, \citenamefont {Vudyasetu},\ and\ \citenamefont
		{Howell}}]{Camacho2009}%
	\BibitemOpen
	\bibfield  {author} {\bibinfo {author} {\bibfnamefont {R.~M.}\ \bibnamefont
			{Camacho}}, \bibinfo {author} {\bibfnamefont {P.~K.}\ \bibnamefont
			{Vudyasetu}},\ and\ \bibinfo {author} {\bibfnamefont {J.~C.}\ \bibnamefont
			{Howell}},\ }\bibfield  {title} {\bibinfo {title} {{Four-wave-mixing stopped
				light in hot atomic rubidium vapour}},\ }\href
	{https://doi.org/10.1038/nphoton.2008.290} {\bibfield  {journal} {\bibinfo
			{journal} {Nature Photonics}\ }\textbf {\bibinfo {volume} {3}},\ \bibinfo
		{pages} {103} (\bibinfo {year} {2009})},\ \Eprint
	{https://arxiv.org/abs/arXiv:1004.3950v1} {arXiv:arXiv:1004.3950v1}
	\BibitemShut {NoStop}%
	\bibitem [{\citenamefont {Albert}\ \emph {et~al.}(2011)\citenamefont {Albert},
		\citenamefont {Dantan},\ and\ \citenamefont {Drewsen}}]{Albert2011}%
	\BibitemOpen
	\bibfield  {author} {\bibinfo {author} {\bibfnamefont {M.}~\bibnamefont
			{Albert}}, \bibinfo {author} {\bibfnamefont {A.}~\bibnamefont {Dantan}},\
		and\ \bibinfo {author} {\bibfnamefont {M.}~\bibnamefont {Drewsen}},\
	}\bibfield  {title} {\bibinfo {title} {{Cavity electromagnetically induced
				transparency and all-optical switching using ion Coulomb crystals}},\ }\href
	{https://doi.org/10.1038/nphoton.2011.214} {\bibfield  {journal} {\bibinfo
			{journal} {Nature Photonics}\ }\textbf {\bibinfo {volume} {5}},\ \bibinfo
		{pages} {633} (\bibinfo {year} {2011})}\BibitemShut {NoStop}%
	\bibitem [{\citenamefont {Cirac}\ \emph {et~al.}(1994)\citenamefont {Cirac},
		\citenamefont {Garay}, \citenamefont {Blatt}, \citenamefont {Parkins},\ and\
		\citenamefont {Zoller}}]{Cirac1994}%
	\BibitemOpen
	\bibfield  {author} {\bibinfo {author} {\bibfnamefont {J.~I.}\ \bibnamefont
			{Cirac}}, \bibinfo {author} {\bibfnamefont {L.~J.}\ \bibnamefont {Garay}},
		\bibinfo {author} {\bibfnamefont {R.}~\bibnamefont {Blatt}}, \bibinfo
		{author} {\bibfnamefont {A.~S.}\ \bibnamefont {Parkins}},\ and\ \bibinfo
		{author} {\bibfnamefont {P.}~\bibnamefont {Zoller}},\ }\bibfield  {title}
	{\bibinfo {title} {Laser cooling of trapped ions: The influence of
			micromotion},\ }\href {https://doi.org/10.1103/PhysRevA.49.421} {\bibfield
		{journal} {\bibinfo  {journal} {Phys. Rev. A}\ }\textbf {\bibinfo {volume}
			{49}},\ \bibinfo {pages} {421} (\bibinfo {year} {1994})}\BibitemShut
	{NoStop}%
	\bibitem [{\citenamefont {Lambrecht}\ \emph {et~al.}(2017)\citenamefont
		{Lambrecht}, \citenamefont {Schmidt}, \citenamefont {Weckesser},
		\citenamefont {Debatin}, \citenamefont {Karpa},\ and\ \citenamefont
		{Schaetz}}]{Lambrecht2017}%
	\BibitemOpen
	\bibfield  {author} {\bibinfo {author} {\bibfnamefont {A.}~\bibnamefont
			{Lambrecht}}, \bibinfo {author} {\bibfnamefont {J.}~\bibnamefont {Schmidt}},
		\bibinfo {author} {\bibfnamefont {P.}~\bibnamefont {Weckesser}}, \bibinfo
		{author} {\bibfnamefont {M.}~\bibnamefont {Debatin}}, \bibinfo {author}
		{\bibfnamefont {L.}~\bibnamefont {Karpa}},\ and\ \bibinfo {author}
		{\bibfnamefont {T.}~\bibnamefont {Schaetz}},\ }\bibfield  {title} {\bibinfo
		{title} {{Long lifetimes and effective isolation of ions in optical and
				electrostatic traps}},\ }\href {https://doi.org/10.1038/s41566-017-0030-2}
	{\bibfield  {journal} {\bibinfo  {journal} {Nat. Photonics}\ }\textbf
		{\bibinfo {volume} {11}},\ \bibinfo {pages} {704} (\bibinfo {year}
		{2017})}\BibitemShut {NoStop}%
	\bibitem [{\citenamefont {Schaetz}(2017)}]{Schaetz2017}%
	\BibitemOpen
	\bibfield  {author} {\bibinfo {author} {\bibfnamefont {T.}~\bibnamefont
			{Schaetz}},\ }\bibfield  {title} {\bibinfo {title} {Trapping ions and atoms
			optically},\ }\href@noop {} {\bibfield  {journal} {\bibinfo  {journal} {J.
				Phys. B: At., Mol. Opt. Phys.}\ }\textbf {\bibinfo {volume} {50}},\ \bibinfo
		{pages} {102001} (\bibinfo {year} {2017})}\BibitemShut {NoStop}%
	\bibitem [{\citenamefont {Schmidt}\ \emph {et~al.}(2018)\citenamefont
		{Schmidt}, \citenamefont {Lambrecht}, \citenamefont {Weckesser},
		\citenamefont {Debatin}, \citenamefont {Karpa},\ and\ \citenamefont
		{Schaetz}}]{Schmidt2018}%
	\BibitemOpen
	\bibfield  {author} {\bibinfo {author} {\bibfnamefont {J.}~\bibnamefont
			{Schmidt}}, \bibinfo {author} {\bibfnamefont {A.}~\bibnamefont {Lambrecht}},
		\bibinfo {author} {\bibfnamefont {P.}~\bibnamefont {Weckesser}}, \bibinfo
		{author} {\bibfnamefont {M.}~\bibnamefont {Debatin}}, \bibinfo {author}
		{\bibfnamefont {L.}~\bibnamefont {Karpa}},\ and\ \bibinfo {author}
		{\bibfnamefont {T.}~\bibnamefont {Schaetz}},\ }\bibfield  {title} {\bibinfo
		{title} {{Optical Trapping of Ion Coulomb Crystals}},\ }\href
	{https://doi.org/10.1103/PhysRevX.8.021028} {\bibfield  {journal} {\bibinfo
			{journal} {Phys. Rev. X}\ }\textbf {\bibinfo {volume} {8}},\ \bibinfo {pages}
		{021028} (\bibinfo {year} {2018})},\ \Eprint
	{https://arxiv.org/abs/1712.08385} {arXiv:1712.08385} \BibitemShut {NoStop}%
	\bibitem [{\citenamefont {Karpa}(2019)}]{Karpa2019}%
	\BibitemOpen
	\bibfield  {author} {\bibinfo {author} {\bibfnamefont {L.}~\bibnamefont
			{Karpa}},\ }\href {https://doi.org/10.1007/978-3-030-27716-1} {\emph
		{\bibinfo {title} {{Trapping single ions and Coulomb crystals with light
					fields}}}},\ SpringerBriefs in Physics\ (\bibinfo  {publisher} {Springer,
		Cham},\ \bibinfo {year} {2019})\BibitemShut {NoStop}%
	\bibitem [{\citenamefont {Schmidt}\ \emph {et~al.}(2020)\citenamefont
		{Schmidt}, \citenamefont {Weckesser}, \citenamefont {Thielemann},
		\citenamefont {Schaetz},\ and\ \citenamefont {Karpa}}]{Schmidt2020}%
	\BibitemOpen
	\bibfield  {author} {\bibinfo {author} {\bibfnamefont {J.}~\bibnamefont
			{Schmidt}}, \bibinfo {author} {\bibfnamefont {P.}~\bibnamefont {Weckesser}},
		\bibinfo {author} {\bibfnamefont {F.}~\bibnamefont {Thielemann}}, \bibinfo
		{author} {\bibfnamefont {T.}~\bibnamefont {Schaetz}},\ and\ \bibinfo {author}
		{\bibfnamefont {L.}~\bibnamefont {Karpa}},\ }\bibfield  {title} {\bibinfo
		{title} {{Optical Traps for Sympathetic Cooling of Ions with Ultracold
				Neutral Atoms}},\ }\href {https://doi.org/10.1103/PhysRevLett.124.053402}
	{\bibfield  {journal} {\bibinfo  {journal} {Phys. Rev. Lett.}\ }\textbf
		{\bibinfo {volume} {124}},\ \bibinfo {pages} {053402} (\bibinfo {year}
		{2020})}\BibitemShut {NoStop}%
\end{thebibliography}
%
%
\end{document}